\documentclass[12pt,a4paper,twoside]{aphs}
\usepackage{amsfonts,amssymb}
\usepackage{graphicx}

 4

\begin{document}

\thispagestyle{empty}




\title{Ag adatoms on Si(111)5$\times$2-Au surface}

\author{A.~ST\c{E}PNIAK}
\address{Max-Planck-Institut f\"{u}r Mikrostrukturphysik, Weinberg 2, 
         06120 Halle, Germany}

\author{M.~KRAWIEC}
\address{M. Curie-Sk\l odowska University, pl. M. Curie-Sk\l odowskiej 1, 
         20-031 Lublin, Poland}

\author{M.~JA\L OCHOWSKI}
\address{M. Curie-Sk\l odowska University, pl. M. Curie-Sk\l odowskiej 1, 
         20-031 Lublin, Poland}

\maketitle


\section{\label{introduction} Introduction}

Gold-induced arrays of atomic chains on flat and vicinal (stepped) Si surfaces 
have attracted considerable interest in recent years due to possibility of
accessing the phenomena characteristic of systems of a reduced dimensionality
\cite{Snijders}. A chain structure on Si(111) surface is known as 5$\times$2-Au
reconstruction \cite{Bishop}-\cite{Bauer}. The unit cell of this reconstruction
is 1.625 nm ($5 \times a_{[1 1 \bar 2}]$) wide and 0.768 nm 
($2 \times a_{[1 \bar 1 0}]$) long. A macroscopically ordered surface with
5$\times$2 reconstruction is observed at gold coverage equal to 0.6 monolayer
(ML) \cite{Barke}. One monolayer refers to the number of silicon atoms in Si 
double layer ($7.83 \times 10^{14}$ atoms/cm$^2$).

The structural model of the Si(111)5$\times$2-Au surface has recently been
proposed by Erwin using density functional theory (DFT) \cite{Erwin}. A
characteristic feature of this model is the Si honeycomb (HC) chain, present
also in Au-induced chain structures on vicinal Si surfaces
\cite{Crain}-\cite{MK_2}. Besides this structure, there are other 
characteristic chains: single- and double-Au chains. In the double-Au chain,
the gold atoms are dimerized, similar as in the case of the Si(553)-Au surface
\cite{MK_2}. Moreover, the Si(111)-5$\times$2-Au surface is decorated by Si
adatoms, which appear in topography measured by scanning tunneling microscopy
(STM) as bright protrusions (PBs) \cite{Yagi}-\cite{Kirakosian}. The BPs tend 
to form a half-filled 5$\times$4 superlattice. This half-filling is reflected 
in parts of the surface with fully occupied 5$\times$4 BP structure and empty 
segments between them, so the average occupation yields one Si adatom per 
5$\times$8 unit cell. 

The Si adatoms play an important role in electronic properties of the surface. 
If the BP coverage is less than one Si adatom per 5$\times$8 unit cell, the 
surface is metallic, while increasing the BP coverage, the surface undergoes a 
metal-insulator transition \cite{Choi}. Furthermore, it was proven 
theoretically that the Si(111)5$\times$2-Au surface is stabilized 
electronically \cite{Erwin,Erwin_2,Stepniak}. It turns out that the system has 
a minimum of energy if doped by electrons equal to one electron per 5$\times$2 
unit cell, that corresponds exactly to a single Si adatom per 5$\times$8 cell.

The scenario of electronic stabilization of the surface was further checked 
with Pb and In adatoms \cite{Stepniak,Stepniak_2}. In the case of Pb atoms, 
they adsorb in the Si-adatom positions, that results in an almost perfect 
5$\times$4 superlattice composed of Pb and Si adatoms. Moreover, the Pb atoms 
modify the electronic properties of the system in the same way as the Si 
adatoms. This was explained in terms of electron doping, since both Si and Pb 
belong to the same group of the periodic table, and have four valence electrons, 
they dope the surface equally \cite{Stepniak}. On the other hand, indium has 
one valence electron less than Pb and Si, thus to dope the surface by the same 
number of electrons as in the case of Pb and Si, we need more In adsorption 
sites. Indeed, it turns out that In has two nonequivalent adsorption sites 
\cite{Stepniak_2}. Similar, as in the case of Pb atoms, one of the adsorption 
sites is located in free Si-adatom positions. The other one is close to the Si 
adatoms, where In atoms form bonds with the Si adatoms. In the STM topography 
images, we deal with three kinds of BPs: the Si adatoms of a native 
Si(111)5$\times$2-Au surface, the In adatoms in positions of the Si adatoms, 
and In atoms bound to the Si adatoms \cite{Stepniak_2}. Interestingly, the 
In-Si structures were visible in the STM topography only at positive sample
bias. In this case, it was not possible to fabricate a perfect 5$\times$4 
adatom superstructures, since In atoms preferred to bond to the Si adatoms 
rather than to fill the empty Si-adatom sites. Such a behavior of Pb and In
adatoms supports the scenario of the electron doping and stabilization of the
Si(111)5$\times$2-Au surface. To shed additional light on the validity of this 
picture, we have chosen different material, namely silver, which has only one 
valence electron. If the above scenario is valid, we should expect more Ag 
adsorption sites, than in the case of Pb and In adatoms. Thus the purpose of 
the present study is to investigate Ag-induced reconstruction on 
Si(111)5$\times$2-Au surface in submonolayer coverage of Ag and verify the 
above assumption.

Using scanning tunneling microscopy together with the first-principles density
functional theory calculations we study structural properties of the
Si(111)5$\times$2-Au surface covered by Ag adatoms. The STM topography data 
show that a submonolayer coverage of Ag does not lead to a well-ordered adatom
chain structure with a periodicity 5$\times$4, as it was observed for Si, Pb and
In adatoms. Instead of that, we observe Ag adatoms located in different
nonequivalent positions within the unit cell, which confirms the electronic
nature of stabilization of the Si(111)5$\times$2-Au surface. Moreover, the DFT
calculations give five different structural models of the
Si(111)5$\times$2-Au/Ag surface. All the models are almost degenerate in energy,
which further supports the above scenario. The rest of the paper is organized as
follows. In Secs. \ref{experiment} and \ref{calculations} details of experiment
and of calculations are provided. The results STM and DFT investigations of the 
Si(111)5$\times$2-Au structure covered by Ag adatoms are presented and 
discussed in Sec. \ref{results}. Section \ref{conclusions} contains the
conclusions summarizing the results.


\section{\label{experiment} Experiment} 

The experimental setup consists of ultrahigh vacuum (UHV) chamber with a base
pressure less than $5 \times 10^{-11}$ mbar equipped with scanning tunneling
microscope (type OmicronVT) and reflection high energy electron diffraction
(RHEED) apparatus. The Si(111) substrate used in the experiment was cut from a 
p-type B-doped sample with a resistivity 0.15 $\Omega$ cm. The sample was
cleaned by flashing several times up to 1500 K. The quality of surface
reconstruction was monitored by RHEED technique. The 5$\times$2 reconstruction
was obtained by deposition of 0.6 ML of Au (in units of a half of Si(111) 
double layer), heating the sample at temperature 1100 K for 60 s and gradually 
cooling down to the room temperature (RT) for 5 min. Next 0.05 ML of Ag was 
evaporated onto the Si(111)5$\times$2-Au surface held at the RT with a rate of 
0.1 ML/min. The quartz-crystal monitor was used to determine both Au and Ag 
coverages. All the STM measurements were performed at the RT.


\section{\label{calculations} Details of calculations}

The calculations have been performed in the local density approximation (LDA)
\cite{Perdew} to DFT, as implemented in the SIESTA code
\cite{Ordejon}-\cite{Artacho_2}. Troullier-Martins norm-conserving 
pseudopotentials \cite{Troullier} have been used in calculations. In the case 
of Au and Ag pseudopotentials the semicore states 5$d$ and 4$d$, respectively, 
as well as the scalar relativistic corrections were included in the
calculations. A basis set in the form of double-$\zeta$ polarized (DZP) 
numerical atomic orbitals  was used for all the atomic species 
\cite{Portal_2,Artacho}. The radii of the orbitals for different species were 
following (in a.u): Au - 7.20 ($5d$), 6.50 ($6s$) and 5.85 ($6p$), Ag - 5.75 
($4d$), 4.71 ($5s$) and 6.20 ($5p$), Si - 7.96 ($3s$), 7.98 ($3p$) and 4.49 
($3d$), and H - 7.55 ($1s$) and 2.94 ($2p$). Eight nonequivalent $k$ points for 
Brillouin zone sampling and a real-space grid equivalent to a plane-wave cutoff 
100 Ry were employed.

The Si(111)5$\times$2-Au/Ag system has been modeled by four Si double layers 
and a vacuum region of 17 \AA. All the atomic positions were relaxed until the
maximum force in any direction was less than 0.04 eV/\AA, except the Si atoms 
in the bottom layer, which were fixed at their bulk ideal positions and
saturated by hydrogen. The lattice constant of Si was fixed at the calculated
value, 5.39 \AA. All the calculations have been performed in the 5$\times$2 
unit cell.


\section{\label{results} Results and discussion}

\begin{figure}[h]
\begin{center}
\includegraphics[width=\linewidth]{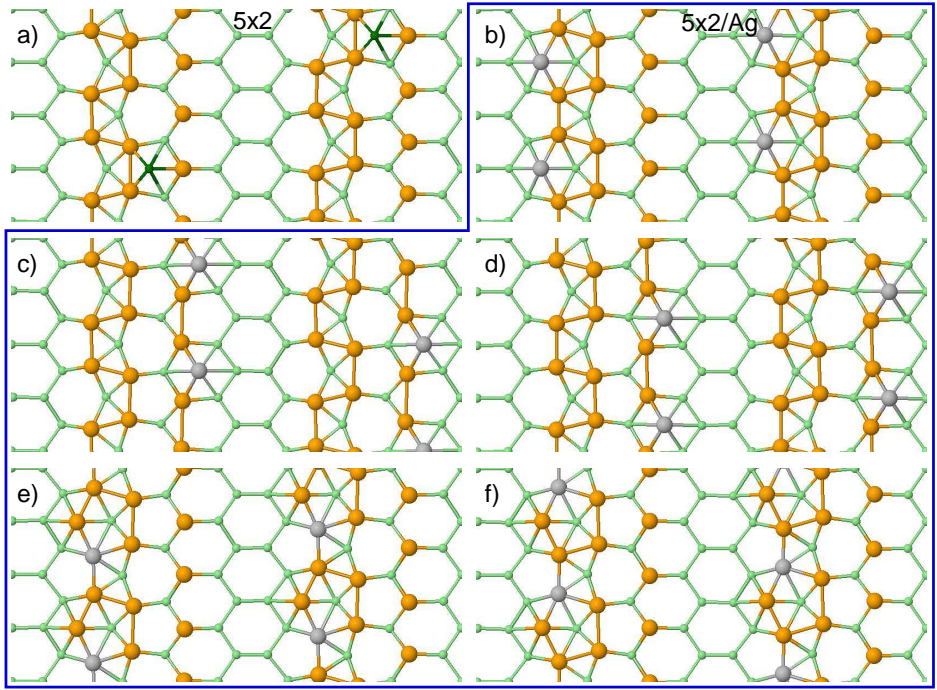}

\vspace{2mm}

\caption{Structural models of the bare [panel a)] and Ag-covered
         Si(111)5$\times$2-Au surface [panels b)-f)]. Different colors reflect
	 different species of the structure: Si atoms - small light-green
	 circles, Au - large orange circles, Ag - large silver circles, and Si
	 adatoms [panel a)] - small dark-green circles. Note, that the model of 
	 the bare surface [panel a)] has 5$\times$4 periodicity due to the 
	 presence of the Si adatoms, while the models of Ag-decorated surface 
	 were calculated in 5$\times$2 unit cell without Si adatoms. 
	 \label{Fig1}}
\end{center}
\end{figure}

The structural model of the Si(111)5$\times$2-Au surface, as proposed by Erwin
\cite{Erwin}, is shown in Fig. \ref{Fig1} a). 
As it was mentioned in Sec. \ref{introduction}, the main building blocks of 
that structure are HC, single- and double-Au chains. Furthermore, the surface 
is decorated by the Si adatoms, forming 5$\times$4 superlattice. Thus in fact 
the structure has 5$\times$4 periodicity. This model, albeit slightly 
simplified, was adapted here to study the adsorption of Ag atoms on the 
Si(111)5$\times$2-Au surface. Namely, the present calculations have been 
limited to the 5$\times$2 unit cell, neglecting the Si adatoms. We expect that 
with this simplification the results do not change significantly, and the 
difference might concern only the number of Ag adsorption sites. In the case of 
the original model, the number of Ag adsorption sites can increase due to 
larger unit cell (5$\times$4) and asymmetry between neighboring 5$\times$2 unit 
cells introduced by Si adatoms.

Five most stable structural models of the Si(111)5$\times$2-Au/Ag surface are
displayed in panels b)-f) of Fig. \ref{Fig1}. As we can see, none of the models 
features Ag adatoms adsorbing in the Si-adatom positions neither in the Pb- nor
in the In-adatom positions. The Si adatoms of the original Si(111)5$\times$2-Au 
surface [panel a)] adsorb between single- and double-Au chains and are bound to 
the Au atoms of both chains. In the case of the double-Au chain, the Si adatoms 
make bonds with dimerized Au atoms. On the other hand, the Ag adatoms are bound 
only to one of the Au chains. The Ag adatoms in model shown in panel b) are 
bound only to the double-Au chain, while the Ag adatoms in models c) and d) - 
to the single-Au chain. The Ag adatoms of model b) are located on the opposite 
side of the double-Au chain than the Si adatoms. Furthermore, the Ag adatoms in
this model are bound to two different Au dimers, rather than to the same 
dimmer, as it observed for Si adatoms. This is reflected in vertical position 
of Ag adatoms. Their $z$-coordinates are by 0.66 \AA\ lower with respect to the 
Si adatoms. It is worthwhile to note that the Ag adatoms in the model b) 
perturb the atomic arrangement of the surface very weakly, and the original 
surface remains almost intact. This is not the case in the remaining models. 
For example, the Ag adatoms bound to the single-Au chain [c) and d)] lead to 
the dimerization of Au atoms in that chain. However, this dimerization is 
weaker than in the double-Au chain. The distance between Au atoms of the same 
dimmer $d_s$ in the single-Au chain is equal to 3.42 \AA, to be compared with 
3.12 \AA\ for the double-Au chain. Similar, the distance between Au atoms of 
different dimmers $d_d$ in single-Au (double-Au) chain yields 4.21 \AA\ 
(4.51 \AA). This yields the dimerization ratio $\delta = d_s/d_d = 0.81$ for 
the single-Au chain, and $\delta = 0.69$ for the double-Au chain. Similar, as 
in model b), the vertical positions of Ag adatoms are lower than that of Si 
adatoms, in this case by 0.41 \AA. In models shown in panels e) and f) of 
Fig. \ref{Fig1}, the modifications of the surface structure are even stronger. 
The Ag atoms substitute the Au atoms of the double chain. As a result, the 
substituted Au atoms are moved to the Ag adsorption sites of model b). However, 
the vertical positions of substituted Au atoms are by 0.26 \AA\ lower than 
that of the Ag adatoms in model b). The lower values of vertical positions of 
Ag adatoms suggest that in STM topography images the apparent height of the Ag 
adatoms should be lower than of the Si adatoms, neglecting differences in 
electronic contributions to the STM signal.

All the models are almost degenerate in energy. The difference between the most 
stable model [panel b)] and the highest-energy model [panel f)] is as small as 
68.9 meV per 5$\times$2 unit cell. This suggests that all five structural 
models should be to an equal extent realized in experiment. Note that In atoms 
adsorb only in two different sites, while in the case of Pb adatoms there was 
only a single adsorption site. Furthermore, as it was mentioned already, none 
of the Ag-adatom positions is equivalent to the Pb nor to In adsorption sites. 
Comparing the number of adsorption sites of Pb, In and Ag atoms, and 
correlating it with corresponding number of valence electrons, we arrive at a 
conclusion that the fewer valence electrons exist in atoms of a given chemical 
element, the more adsorption sites available. Thus, this is a strong argument 
supporting the scenario of electronic stabilization of the surface. 

\begin{figure}[h]
\begin{center}
\includegraphics[width=\linewidth]{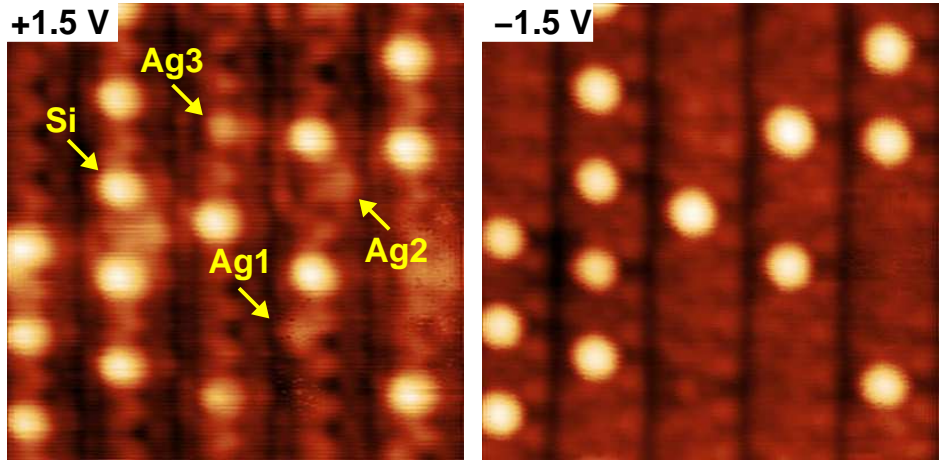}

\vspace{2mm}

\caption{$8 \times 8$ nm$^2$ STM topography of the same area of the
          Si(111)5$\times$2-Au surface covered by 0.05 ML of Ag. The data were
	  acquired at tunneling current $I = 0.3$ nA and bias voltages: 
	  $U = + 1.5$ V (left panel) and $U = - 1.2$ V (right panel). Note
	  different kinds of bright protrusions. The arrows point to different
	  kinds of bright protrusions. Si indicates one of the Si adatoms, 
	  while Ag1 to Ag3 - Ag adatoms located in three nonequivalent 
	  adsorption sites. \label{Fig2}}
\end{center}
\end{figure}

To verify the above theoretical predictions, we have performed the STM
measurements of the Si(111)5$\times$2-Au surface covered by 0.05 ML of Ag. The 
STM topography images are shown in Fig. \ref{Fig2}.
First of all, the Si adatoms are much better visible than the Ag adatoms 
(Ag1 to Ag3), confirming the DFT predictions. Furthermore, the Ag adatoms 
are located in different adsorption sites. Three nonequivalent sites are marked
by Ag1 to Ag3. Comparing the positions of the Ag adatoms to the positions
obtained by DFT calculations (Fig. \ref{Fig1}), we can conclude that Ag1
corresponds either to the Ag adatom in the model b) or Au adatom in model e) or 
f). Unfortunately, further verification is impossible on the basis of the STM
topography data, since the calculated positions of the Ag and Au adatoms are
equivalent, and the apparent heights of all the Ag and Au adatoms are 
comparable. Note that in the case of In adatoms it was possible to distinguish 
them in scanning tunneling spectroscopy (STS) \cite{Stepniak_2}. Here, however,
the calculated densities of states (DOS) show rather small variations for 
adatoms in models b), e) and f). Thus the differences in DOS may not be 
resolved in the STS characteristics. A similar comparison between Fig. 
\ref{Fig1} and Fig. \ref{Fig2} indicates that the Ag2 adatom is the realization 
of the model c) or d). For the same reasons, a more detailed analysis is very 
tricky at present. Finally, we also observe Ag3 adatom located in the position,
which is not reproduced in any of the models displayed in Fig. \ref{Fig1}. The
reason for that is likely due to neglecting the Si adatoms in present 
calculations. We expect this adsorption site to appear in calculations in 
complete Si(111)5$\times$2-Au model of Ref. \cite{Erwin}, as was discussed at
the begging of this section.

The above STM topography results confirm the DFT predictions that the number of 
Ag adsorption sites should be higher than in the case of In and Pb adatoms. As
it was discussed previously, such findings are required by the scenario of 
electronic stabilization of the Si(111)5$\times$2-Au surface. Thus the behavior 
of Si, Pb, In and Ag adatoms altogether unambiguously proofs the realization of 
this scenario.


\section{\label{conclusions} Conclusions}

In conclusion, we have performed the STM and DFT study of the structural
properties of the Si(111)5$\times$2-Au surface covered by Ag adatoms. The DFT
results predict the existence of five different adsorption sites. None of them 
is equivalent to any of the Si, Pb and In adsorption sites. The DFT predictions 
have been confirmed by the STM topography images, in which three different
adsorption sites were revealed. Two of them are reproduced well by the present 
calculations. Furthermore, the Ag adatoms appear to be lower that the Si
adatoms in the STM data, that allows for a clear distinction between them. The
obtained results point to a significance of the electron doping and entirely 
confirm the picture of the electronic stabilization of the Si(111)5$\times$2-Au 
surface


\subsection*{Acknowledgements}

This work has been supported by the Polish Ministry of Education and Science
under Grant No. N N202 330939.



\end{document}